# Full quantum analysis of complete population transfer using frequency boost

Fatemeh Ahmadinouri [1], Mehdi Hosseini[1a] and Farrokh Sarreshtedari [2]

*[1]Department of physics, Shiraz University of Technology, Shiraz, 313-71555, Iran*

*[2]Magnetic Resonance Research Laboratory, Department of Physics, University of*

*Tehran, Tehran 143-9955961, Iran*

In this paper, we have proposed and demonstrated a new method of atomic population transfer. Transition dynamic of a two-level system is studied in a full quantum description of the Jaynes-Cummings model. Solving the time-dependent Schrödinger equation, we have investigated the transition probabilities numerically and analytically by using a sudden boost of the laser frequency. The results show that complete population transfer can be achieved by adjusting the time of the frequency boost.

## I. INTRODUCTION

Population transfer of quantum states play a fundamental important role in variety of fields of physics [1–5]. For this reason, finding an efficient transferring mechanism is an important issue in atomic physics.

One of the fundamental schemes for full quantum modeling of the system dynamics is the *Jaynes-Cummings* (JC) model [6–8]. JC model explains analytically the interaction of a quantized electromagnetic field with a two-level system. The early history of this model returns back to magnetic resonance [9,10] and it was presented in 1963 by Jaynes and Cummings [11]. The JC model is experimentally realized with the high-Q superconducting cavities and a Rydberg atom [12]. This scheme describes several interesting phenomena such as collapse and revivals [13], atom-field entanglement [12,14], squeezing [15–17] and Rabi Oscillation (RO) [8].

---

[a] hosseini@sutech.ac.ir

Rabi Oscillation was first introduced in 1937 and explained by interaction of the oscillating magnetic field and the magnetic moments [18]. The Rabi oscillations mainly describe the interaction of the two-level atom with a radiation field [19]. Such behavior is observed in the different fields such as quantum dots [20,21], trapped ions [22,23], superconducting quantum devices [24], semiconductors [25], surface plasmons [26], Bose–Einstein condensates [27], diamond nitrogen-vacancy centers [28] and Josephson junction qubits [29].

Here, we have numerically and analytically studied the population transfer in a two-level system using the *Jaynes-Cummings* model. In this method, the variation of Rabi Oscillations and transition probabilities are investigated using sudden frequency change and it is shown that by tuning the time of frequency boost, the complete population transfer can be obtained.

## II. THEORETICAL MODEL

In this work, the transition probabilities have been studied using a full quantum model of the Jaynes-Cummings (JC). When an external electric field interacts with a two-level system, the JC Hamiltonian (Eq. 1) includes three terms for the atom, field, and interaction between them respectively as follow [30]:

$$H = -\frac{1}{2}\hbar\omega_0\sigma_z + \hbar\omega_L(a^\dagger a + \frac{1}{2}) + \hbar\omega_1(a\sigma_+ + a^\dagger\sigma_-) \tag{1}$$

Where ℏ is Planck constant, $\omega_0$ is the atomic transition frequency, $\omega_L$ is the laser frequency. As well as $a$, $a^\dagger$ are annihilation and creation operators respectively, and $\sigma_-$, $\sigma_+$, $\sigma_z$ are Pauli spin matrices. In this definition, $\omega_1$ describes the coupling strength between atom and field which is determined by the atomic transition dipole moment and the laser field and is proportional to Rabi frequency (Eq. 2) [30]:

$$\Omega = \sqrt{(\omega_L - \omega_0)^2 + (2\omega_1)^2} \qquad (2)$$

By acting the JC Hamiltonian (Eq. 1) on eigenstates |g, n⟩ (ground state, n photons) and |e, n-1⟩ (excited state, n photons) (Eq. 3), the final Hamiltonian (Eq. 4) can be obtained [30].

$$H_{atom}\left|g;n\right\rangle = -\frac{1}{2}\hbar\omega_0\left|g;n\right\rangle \qquad H_{atom}\left|e;n-1\right\rangle = \frac{1}{2}\hbar\omega_0\left|e;n-1\right\rangle$$

$$H_{field}\left|g;n\right\rangle = \hbar\omega_L(n+\frac{1}{2})\left|g;n\right\rangle \qquad H_{field}\left|e;n-1\right\rangle = \hbar\omega_L(n-\frac{1}{2})\left|e;n-1\right\rangle \qquad (3)$$

$$H_{\text{int}}\left|g;n\right\rangle = \hbar\omega_1\sqrt{n}\left|e;n-1\right\rangle \qquad H_{\text{int}}\left|e;n-1\right\rangle = \hbar\omega_1\sqrt{n}\left|g;n\right\rangle$$

Finally, by using this Hamiltonian (Eq. 4) and by numerically solving the time-dependent Schrödinger equation the transition probabilities are obtained using Runge–Kutta method. In this calculation, $n$ is equal to one.

$$H = \hbar \begin{pmatrix} (n+\frac{1}{2})\omega_L - \dfrac{\omega_0}{2} & \omega_1\sqrt{n} \\ \omega_1\sqrt{n} & (n-\frac{1}{2})\omega_L + \dfrac{\omega_0}{2} \end{pmatrix} \qquad (4)$$

It is worth mentioning that for the simplicity, all the parameters are dimensionless by the relation (5) in our calculations.

$$\tilde{\omega}_0 = 1, \tilde{t} = t\,\omega_0, \tilde{\omega}_1 = \frac{\omega_1}{\omega_0}, \tilde{\omega}_L = \frac{\omega_L}{\omega_0} \qquad (5)$$

## III. RESULTS AND DISCUSSION

As mentioned before, when a two-level atom interacts with a resonant external field, the system population oscillates between the ground and excited states. Here, the system population transfer and variation of Rabi-oscillations have been investigated by a sudden change in laser frequency towards achieving the stable and complete population transfer.

In section A, the effect of a sudden increase of the laser frequency on the transition probabilities has been investigated. In section B, the transition probabilities using two frequency boosts have been studied. Moreover, it is shown that the analytical solution of transition probabilities is in agreement with the numerical solution in section C.

### A. A sudden increase of the laser frequency

In Fig. 1, the transition probabilities versus time are depicted for $\tilde{\omega}_l$=0.1. The initial and final normalized laser frequency is considered $\tilde{\omega}_a$ and $\tilde{\omega}_b$, respectively. In order to investigate the effect of the sudden increase of the laser frequency, the time of the frequency change is set when the probability of the initial ROs is minimum and maximum in the left and right column in Fig. 1, respectively. Also in the middle column, this frequency increase is tuned when the probability of the initial ROs is in half of the maximum. This figure reveals that by sudden increase in the laser frequency i) the amplitude of the probability oscillation is intensively decreased and whereby the final probability stability is enhanced. ii) The probability of system population transfer remains constant at the same probability of the frequency boost time. Therefore, it is possible to achieve the desired probability by adjusting the time of the sudden increase of the laser frequency. Indeed, the boosting time plays a key role in the final probability. So, in order to achieve the maximum possible probability, the time of the frequency boost is set to the time that the probability of the initial ROs is maximum (right column). Accordingly, the complete transition (Fig. 1 (f)) is observed when the initial ROs oscillate between zero and one (middle row). It means that the initial frequency is tuned to the atomic transition frequency ($\tilde{\omega}_a$=1). This figure also shows that by decreasing (upper row) or increasing (lower row) the initial frequency, probabilities are reduced.

In Ref. [20] it has been shown that by applying the chirped laser source in a limited time interval, the population of the states can be tuned which the results are similar to the method of the frequency boost that is proposed here.

The transition probability versus time for different final laser frequency is illustrated in Fig. 2, in which the initial laser frequency is equal to the atomic transition frequency. The time of the frequency boost is adjusted when the probability of the initial ROs is half of the maximum (0.5). This figure shows that the amplitude of the secondary ROs is reduced by increasing the final laser frequency. Therefore, by increasing the frequency differences, stable population transfer with the lowest oscillation can be achieved.

In Fig. 3 transition probability versus time is depicted for different coupling strengths while $\tilde{\omega}_a$=1, $\tilde{\omega}_b$=5. The time of the frequency changing is tuned to the complete transition probability. As the ROs frequency increases by increasing the coupling strength, the frequency should change more abruptly to achieve the complete population transfer. In other words, by decreasing the coupling strength, the population transfer would be insensitive to the small variations of the frequency boost time and thus the robustness increases.

It can be inferred that in the sudden frequency boost method, the robust and complete population transfer would be achieved by adjusting the time of the frequency changing, the value of frequency changing and coupling strength.

**B. Robust population transfer using two frequency boosts**

So far, the population can be only transferred to the excited state by adjusting system parameters. In this section, the population transfer is investigated in such a way that ensures that the system has been in the ground state for a long time ago. Then it tries to transfer this population to the excited state. To achieve this goal, the two sudden frequency boosts are proposed. The transition probability versus time for $\tilde{\omega}_1=0.1$ is depicted in Fig. 4. The initial, second and final laser frequencies are considered as $\tilde{\omega}_a$, $\tilde{\omega}_b$, and $\tilde{\omega}_c$, respectively. As it can be seen, $\tilde{\omega}_b$ is tuned to the atomic transition frequency. All population is in the ground state if the time of first sudden changing of the laser frequency is set to when the probability of the second ROs is minimum. After that by adjusting the final frequency boost to when the probability of the second ROs is maximum, the stable and complete population transfer occurs.

**C. Analytical solution**

In this section, by analytical solution, the stable and full population transfer to the excited state is investigated. For this aim, the eigenvalue of JC Hamiltonian (Eq. 6) is calculated (Eq. 7).

$$H = \hbar\omega_0 \begin{pmatrix} \frac{1}{2}(3\tilde{\omega}_L - 1) & \tilde{\omega}_1 \\ \tilde{\omega}_1 & \frac{1}{2}(\tilde{\omega}_L + 1) \end{pmatrix} \tag{6}$$

$$\lambda = \hbar\omega_0(\tilde{\omega}_L \mp \frac{\tilde{\Omega}}{2}) \tag{7}$$

Then by solving the time-dependent Schrödinger equation, the wave function is achieved (Eq. 8).

$$i\hbar\frac{\partial\psi}{\partial t} = H\psi \rightarrow \psi_t = e^{\frac{H}{i\hbar}(t-t_0)}\psi_0 \tag{8}$$

Where $\Psi_0$ is the wave function in the initial condition, $t_0$ is the initial time, $\Psi_t$ and $t$ are the final wave function and the final time, respectively.

To find the final wave function, we rewrite the Hamiltonian in the diagonal form (Eq. 9). The $d$ index expresses that the Hamiltonian is diagonal. In this solution, $R$ is the operator transformation (Eq. 11) which is obtained by Eq. 10. In this relation, $\Psi_1(t)$ and $\Psi_2(t)$ are the wave functions in the ground and excited state, respectively.

$$\psi_t = (R^{-1} e^{\frac{H_d(t-t_0)}{i\hbar}} R)\psi_0 \tag{9}$$

$$H \begin{pmatrix} \psi_1(t) \\ \psi_2(t) \end{pmatrix} = \lambda \begin{pmatrix} \psi_1(t) \\ \psi_2(t) \end{pmatrix} \tag{10}$$

$$R = \begin{pmatrix} 2\tilde{\omega}_1 & -(\tilde{\omega}_L + \tilde{\Omega} - 1) \\ -2\tilde{\omega}_1 & (\tilde{\omega}_L - \tilde{\Omega} - 1) \end{pmatrix} \tag{11}$$

Finally, by solving the Schrödinger equation (Eq. 9), the wave function (Eq. 12) and so the final transition probability can be achieved. In this calculation, $\tilde{t}_{bst}$ is the time of the frequency boost. In relation (12), the wave function is obtained before (a) and after (b) of the frequency variation. The parameters of this relation are shown in Eq. 13. In this work, it is assumed that the system has been in the ground state for the initial time.

$$\begin{cases} a) \tilde{t} < \tilde{t}_{bst} : \begin{pmatrix} \psi_{1a}(\tilde{t}) \\ \psi_{2a}(\tilde{t}) \end{pmatrix} = \frac{1}{\left| -4\tilde{\omega}_1 \tilde{\Omega} \right|} \begin{pmatrix} C_{\tilde{\omega}_a}(\tilde{t}) & D_{\tilde{\omega}_a}(\tilde{t}) \\ E_{\tilde{\omega}_a}(\tilde{t}) & F_{\tilde{\omega}_a}(\tilde{t}) \end{pmatrix} \begin{pmatrix} 1 \\ 0 \end{pmatrix} = \frac{1}{\left| -4\tilde{\omega}_1 \tilde{\Omega} \right|} \begin{pmatrix} C_{\tilde{\omega}_a}(\tilde{t}) \\ E_{\tilde{\omega}_a}(\tilde{t}) \end{pmatrix} \\[4mm] b) \tilde{t} \geq \tilde{t}_{bst} : \begin{pmatrix} \psi_{1b}(\tilde{t}) \\ \psi_{2b}(\tilde{t}) \end{pmatrix} = \frac{1}{\left| -4\tilde{\omega}_1 \tilde{\Omega} \right|} \begin{pmatrix} C_{\tilde{\omega}_b}(\tilde{t}) & D_{\tilde{\omega}_b}(\tilde{t}) \\ E_{\tilde{\omega}_b}(\tilde{t}) & F_{\tilde{\omega}_b}(\tilde{t}) \end{pmatrix} \begin{pmatrix} C_{\tilde{\omega}_a}(\tilde{t}_{bst}) \\ E_{\tilde{\omega}_a}(\tilde{t}_{bst}) \end{pmatrix} = \frac{1}{\left| -4\tilde{\omega}_1 \tilde{\Omega} \right|} \begin{pmatrix} C_{\tilde{\omega}_a}(\tilde{t}_{bst})C_{\tilde{\omega}_b}(\tilde{t}) + E_{\tilde{\omega}_a}(\tilde{t}_{bst})D_{\tilde{\omega}_b}(\tilde{t}) \\ C_{\tilde{\omega}_a}(\tilde{t}_{bst})E_{\tilde{\omega}_b}(\tilde{t}) + E_{\tilde{\omega}_a}(\tilde{t}_{bst})F_{\tilde{\omega}_b}(\tilde{t}) \end{pmatrix} \end{cases}$$ (12)

$$\begin{cases} C_{\tilde{\omega}_L}(\tilde{t}) = 2\tilde{\omega}_1(\hat{\omega}_L - \tilde{\Omega} - 1)e^{-i(\tilde{\omega}_L - \frac{\tilde{\Omega}}{2})(\tilde{t} - \tilde{t}_0)} - 2\tilde{\omega}_1(\hat{\omega}_L + \tilde{\Omega} - 1)e^{-i(\tilde{\omega}_L + \frac{\tilde{\Omega}}{2})(\tilde{t} - \tilde{t}_0)} \\[2mm] D_{\tilde{\omega}_L}(\tilde{t}) = (\tilde{\omega}_L - \tilde{\Omega} - 1)(\tilde{\omega}_L + \tilde{\Omega} - 1)e^{-i(\tilde{\omega}_L + \frac{\tilde{\Omega}}{2})(\tilde{t} - \tilde{t}_0)} - (\tilde{\omega}_L - \tilde{\Omega} - 1)(\tilde{\omega}_L + \tilde{\Omega} - 1)e^{-i(\tilde{\omega}_L - \frac{\tilde{\Omega}}{2})(\tilde{t} - \tilde{t}_0)} \\[2mm] E_{\tilde{\omega}_L}(\tilde{t}) = 4\tilde{\omega}_1^2(e^{-i(\tilde{\omega}_L - \frac{\tilde{\Omega}}{2})(\tilde{t} - \tilde{t}_0)} - e^{-i(\tilde{\omega}_L + \frac{\tilde{\Omega}}{2})(\tilde{t} - \tilde{t}_0)}) \\[2mm] F_{\tilde{\omega}_L}(\tilde{t}) = 2\tilde{\omega}_1(\hat{\omega}_L - \tilde{\Omega} - 1)e^{-i(\tilde{\omega}_L + \frac{\tilde{\Omega}}{2})(\tilde{t} - \tilde{t}_0)} - 2\tilde{\omega}_1(\hat{\omega}_L + \tilde{\Omega} - 1)e^{-i(\tilde{\omega}_L - \frac{\tilde{\Omega}}{2})(\tilde{t} - \tilde{t}_0)} \end{cases}$$ (13)

In Fig. 5 the transition probabilities versus time are illustrated for numerical and analytical Solutions, $\tilde{\omega}_1$=0.1, $\tilde{\omega}_a$=1, and $\tilde{\omega}_b$=5. It can be seen that analytical calculation is confirmed by the numerical results.

## IV. CONCLUSION

In this work, the population transfer in a two-level system has been numerically and analytically studied for a sudden boost of laser frequency. The population transfer is investigated by numerically solving of the Jaynes-Cummings Hamiltonian. This method reveals that by a sudden increase in the laser frequency, the amplitude of the probability oscillation is intensively decreased. In addition, the system probability remains constant at the same probability of the frequency boost time. Therefore, by adjusting the coupling strength, the value and the time of the laser frequency boost, complete population transfer can be achieved. It has also shown that the numerical results are in agreement with the analytical calculations.

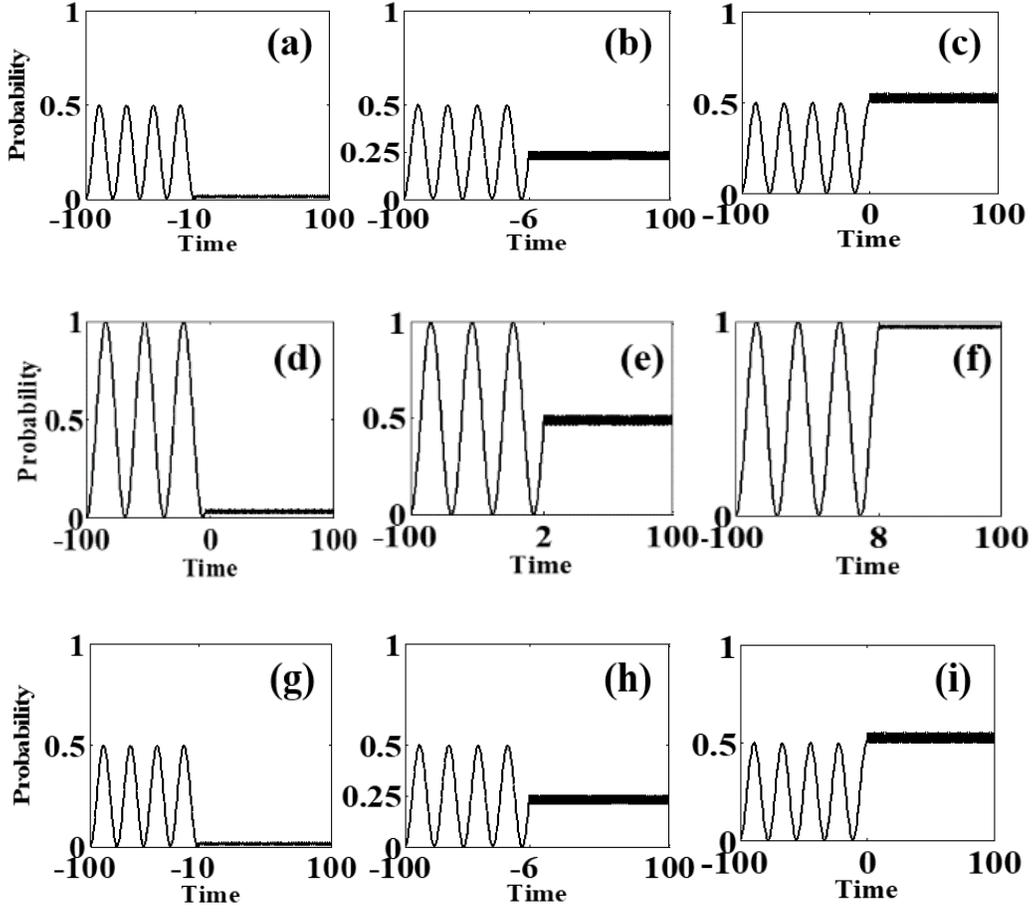

FIG. 1. Transition probability versus time $-100 < \tilde{t} < 100$, $\tilde{\omega}_I = 0.1$ (a), (b), (c): $\tilde{\omega}_a = 0.8$, $\tilde{\omega}_b = 5$ (d),(e), (f): $\tilde{\omega}_a = 1$, $\tilde{\omega}_b = 5$ (g), (h), (i): $\tilde{\omega}_a = 1.2$, $\tilde{\omega}_b = 5$

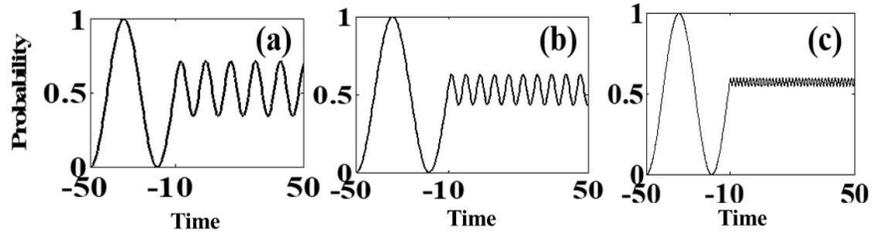

FIG. 2. Transition probability versus time $-50 < \tilde{t} < 50$, $\tilde{\omega}_1 = 0.1$, $\tilde{\omega}_a = 1$

(a) $\tilde{\omega}_b = 1.5$ (b) $\tilde{\omega}_b = 2$ (c) $\tilde{\omega}_b = 5$

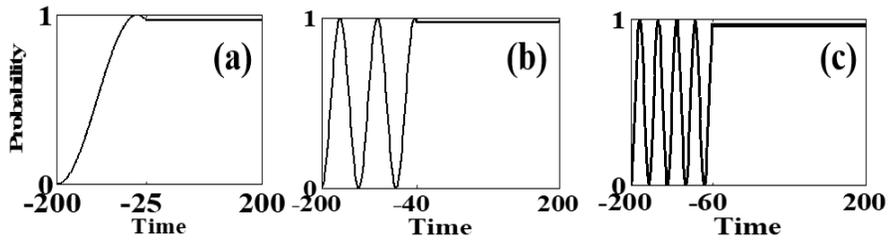

FIG. 3. Transition probability versus time $-200 < \tilde{t} < 200$, $\tilde{\omega}a = 1$, $\tilde{\omega}_b = 5$ (a)

$\tilde{\omega} = 0.01$ (b) $\tilde{\omega}_1 = 0.05$ (c) $\tilde{\omega}_1 = 0.1$

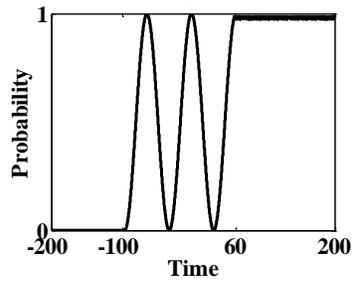

FIG. 4. Transition probability versus time $-200 <$

$\tilde{t} < 200$, $\tilde{\omega}_1 = 0.1$, $\tilde{\omega}_a = 3$, $\tilde{\omega}_b = 1$, $\tilde{\omega}_c = 3$

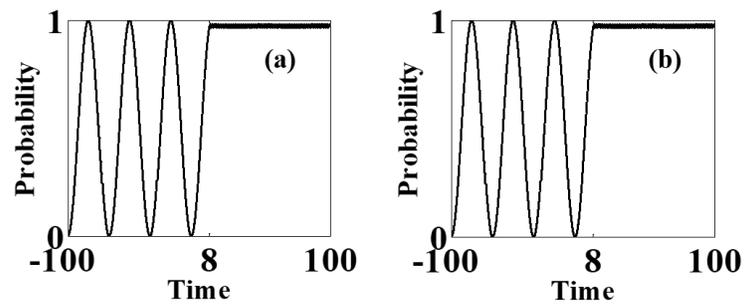

FIG. 5. Transition probability versus time $-100 < \tilde{t} < 100$, $\tilde{\omega}_1 = 0.1$, $\tilde{\omega}_a = 1$, $\tilde{\omega}_b = 5$ (a) numerical solution (b) analytical Solution